\documentclass[11pt,a4paper]{article}
\usepackage[T1]{fontenc}
\usepackage[utf8]{inputenc}
\IfFileExists{lmodern.sty}{\usepackage{lmodern}}{}
\usepackage[margin=2.4cm]{geometry}
\usepackage{amsmath,amssymb,mathtools,bm}
\numberwithin{equation}{section}
\usepackage[expansion=false]{microtype}
\usepackage[hidelinks]{hyperref}
\newcommand{\ii}{\mathrm i}
\newcommand{\ee}{\mathrm e}
\newcommand{\dd}{\mathrm d}
\newcommand{\ket}[1]{\lvert#1\rangle}
\newcommand{\bra}[1]{\langle#1\rvert}
\newcommand{\norm}[1]{\lVert#1\rVert}
\newcommand{\Ntot}{\mathcal N}
\newcommand{\DelE}{\Delta E}
\newcommand{\ww}{\omega_0}
\newcommand{\mut}{\mu_{\rm tr}}
\newcommand{\cl}{c}
\newcommand{\wtr}{\omega_{\rm tr}}
\newcommand{\Ec}{E_{\rm c}}
\newcommand{\xr}{x_{\rm red}}

\AtBeginDocument{%
  \setcounter{secnumdepth}{2}%
  \expandafter\def\csname p@subsection\endcsname{}%
}
\hypersetup{pdftitle={Where should we look for the deviation from quantum mechanics?},
pdfauthor={Jérôme Nicolas and Albert Schwarz}}
\begin{document}
\title{Where should we look for the deviation from quantum mechanics?}
\author{Jérôme Nicolas\thanks{\href{mailto:jerome.nicolas@ac-aix-marseille.fr}{\texttt{jerome.nicolas@ac-aix-marseille.fr}}}\\
{\small Lycée Saint Charles, Marseille, France}\\[0.8em]
Albert Schwarz\thanks{Contact author: \href{mailto:schwarz@math.ucdavis.edu}{\texttt{schwarz@math.ucdavis.edu}}}\\
{\small Department of Mathematics, University of California,}\\
{\small Davis, CA 95616, USA}}
\date{}
\maketitle

\begin{abstract}
We argue that the experimental measurement of the lifetime and decay law of
metastable helium is a promising way to find a deviation from quantum mechanics.
We analyze the decay of the metastable state for a quartic deformation of quantum mechanics. We find that the absolute change of the mean lifetime can be significant even if the modification is small. We show that, in the Wigner--Weisskopf approximation, the deviation from the exponential law in the modified theory can occur on time scales comparable with the lifetime.
\end{abstract}

\section{Introduction}\label{sec:intro}

It is natural to verify the predictions of quantum mechanics (QM) at the level of elementary
particles or atoms by analyzing their energy levels, scattering or decay. If the deviation from quantum
mechanics is small, then the shift of energy levels should also be small; the modification of scattering may be even smaller because the
collision time is short. This suggests studying the decay of almost stable particles
or atoms. We can arrive at the same conclusion by noticing that the relative change
of the width $|\delta\Gamma|/\Gamma$ is the same as the relative change
of the lifetime $|\delta\tau|/\tau$, to first order in $\delta\Gamma$. (Here $\Gamma$ stands for the decay rate (width),
$\tau=1/\Gamma$ denotes the lifetime, and $\delta$ denotes a small change.)
One should expect that
$\delta\Gamma$ is small, but even in this case $|\delta\tau|$ can be large for long-lived
particles. It seems that a good chance to find a violation of the rules of quantum mechanics
is to analyze the lifetime and decay law of metastable helium experimentally. The choice of
this atom is prompted by the fact that its lifetime is really long (about two hours) and by
the fact that the lifetime can be calculated theoretically (for more complicated atoms it
is very difficult to calculate the lifetime precisely). The lifetime of metastable helium
was studied in~\cite{H}. The measured value, $7870(510)$~s, has a relative uncertainty of
about $6.5\%$ and agrees with the theoretical prediction~\cite{LP} within this uncertainty. A more precise experiment can either confirm
QM within experimental accuracy or show some deviation from QM.

We study the decay of the metastable state of helium in modified quantum mechanics. The
modification can be constructed either in the framework of Weinberg's nonlinear quantum
mechanics~\cite{W} or in the framework of the paper~\cite{S}, where quantum mechanics was
interpreted as classical mechanics close to equilibrium. The use of~\cite{S} leads to
simpler calculations. We will follow this paper. (Notice, however, that Weinberg's
nonlinear quantum mechanics leads qualitatively to the same results with an appropriate choice of modification of QM, see the \hyperlink{weinberg-appendix}{Appendix}.) It follows from our calculations in Section~\ref{sec:quartic} that the absolute change in the mean lifetime can be appreciable even in the case when the modification
is small. Another conclusion of Section~\ref{sec:quartic}: generically the decay law is nonexponential on the lifetime
scale in the Wigner--Weisskopf approximation, whereas the decay law of ordinary quantum mechanics is usually close to an exponential law everywhere except some intervals at the beginning and the end of the process.
As we emphasized, metastable helium is a long-lived atom for which the lifetime can be
calculated precisely from first principles. For long-lived atoms whose lifetime cannot be
calculated precisely, finding the lifetime experimentally cannot by itself provide a
test of quantum mechanics. However, an unexpected deviation from the exponential law can
serve as an indication of a violation of the laws of quantum mechanics.

\section{Lifetime of metastable helium}\label{sec:ordinary}

In this section we present a well-known calculation of the lifetime of metastable helium in the form appropriate for generalization.
Field conventions and technical details are given in the Appendix.

Let us consider helium as a two-level quantum system. The lower level is the
nondegenerate state $\ket0=\ket{1\,{}^1S_0}$, of energy $E_0$, and the upper
level is the $2\,{}^3S_1$ triplet, of energy $E_1$. The total electronic angular
momentum quantum number is $J=1$ for this triplet. We use a Cartesian basis
$\ket i$, $i=1,2,3$, transforming as the components of a vector under rotations.
Let $\hbar$ be the reduced Planck constant. Define the positive energy gap
$\DelE=E_1-E_0$ and angular frequency $\ww=\DelE/\hbar$.

With a suitable choice of phases, rotational invariance gives
$\langle0|\mu_j|i\rangle=\mut\delta_{ji}$,
where $\mu_j$ are the Cartesian components of the atomic magnetic-moment operator,
$\delta_{ji}$ is the Kronecker symbol and $\mut$ is a real transition coefficient.

The transition part of the interaction at the position $\mathbf0$ of the atom is
\begin{equation}\label{eq:Hint}
 H_{\rm int}=-\mut\sum_{i=1}^3
       (\ket0\bra i+\ket i\bra0)B_i(\mathbf0).
\end{equation}
Here $B_i$ are the components of the quantized magnetic field. For real
metastable helium, $\mut$ is an effective transition magnetic moment.
A photon mode $\alpha=(\mathbf k,\lambda)$ has wave vector $\mathbf k$,
polarization $\lambda=1,2$ and angular frequency
$\omega_\alpha=\cl|\mathbf k|$, where $\cl$ is the speed of light.
Every sum over $\alpha$ runs over all such modes.
Let $\ket{0_\gamma}$ be the photon vacuum and
$\ket{1_\alpha}$ the corresponding one-photon state.
For emission, define
\begin{equation}\label{eq:g}
 g_{i\alpha}=\langle0;1_\alpha|H_{\rm int}|i;0_\gamma\rangle.
\end{equation}

We retain the terms in~\eqref{eq:Hint} that conserve the number of atomic
excitations plus photons. This is the rotating-wave approximation (RWA),
which defines the effective model used below.

Take an initial normalized upper-state polarization
$\ket\chi=\sum_{i=1}^{3} u_i\ket i$, with $\sum_{i=1}^{3}|u_i|^2=1$, and no photons.
The state of one atom and the field in the one-excitation sector is
\begin{equation}\label{eq:state}
 Z(t)=\sum_{i=1}^{3}\psi_i(t)\ket{i;0_\gamma}
      +\sum_\alpha\phi_\alpha(t)\ket{0;1_\alpha}.
\end{equation}

We use a normalized joint state. The initial condition is
$\psi_i(0)=u_i$, $\phi_\alpha(0)=0$.
We write $x(t)=n_1(t)$ where $n_1=\sum_{i=1}^{3}|\psi_i|^2.$

We introduce the notation
\begin{equation}\label{eq:freephase}
 \psi_i(t)=a_i(t)\ee^{-\ii E_1t/\hbar},\qquad
 \phi_\alpha(t)=b_\alpha(t)\ee^{-\ii(E_0+\hbar\omega_\alpha)t/\hbar}.
\end{equation}

In this notation, the Schr\"odinger equations~\eqref{eq:ordinary-psi}--\eqref{eq:ordinary-phi} of the Appendix give
\begin{align}
 \dot a_i(t)&=-\frac{\ii}{\hbar}\sum_\alpha
       \bar g_{i\alpha}b_\alpha(t)\ee^{\ii(\ww-\omega_\alpha)t},\label{eq:ordinary-a}\\
 \dot b_\alpha(t)&=-\frac{\ii}{\hbar}\sum_{j=1}^{3}
       g_{j\alpha}a_j(t)\ee^{-\ii(\ww-\omega_\alpha)t}.\label{eq:ordinary-b}
\end{align}

Eliminating $b_\alpha(t)$ yields the memory equation
\begin{equation}\label{eq:memory-matrix}
 \begin{aligned}
 \dot a_i(t)&=-\sum_{j=1}^{3}\int_0^tK_{ij}(t-t')a_j(t')\,\dd t',\\
 K_{ij}(s)&=\frac1{\hbar^2}\sum_\alpha
    \bar g_{i\alpha}g_{j\alpha}\ee^{\ii(\ww-\omega_\alpha)s}.
 \end{aligned}
\end{equation}

Rotational invariance in the continuum limit implies
$K_{ij}(s)=\delta_{ij}K(s)$. Hence
$a_i(t)=u_i a(t)$, and
\begin{equation}\label{eq:ordinary-memory}
 \dot a(t)=-\int_0^tK(s)a(t-s)\,\dd s,\qquad a(0)=1.
\end{equation}

The expression $x(t)=n_1(t)=|a(t)|^2$ can be interpreted as the probability
that the atom is still in the metastable state at time $t$.

 For any fixed $i$, define
the spectral density
\begin{equation}\label{eq:J}
 \begin{aligned}
 J(\omega)&=\frac1{\hbar^2}\sum_\alpha|g_{i\alpha}|^2
             \delta(\omega-\omega_\alpha),\\
 K(s)&=\int_0^\infty J(\omega)\ee^{\ii(\ww-\omega)s}\,\dd\omega{,}
 \end{aligned}
\end{equation}
where $\delta$ stands for the Dirac delta function.

The function $J(\omega)$ is independent of $i$ in the isotropic limit.
We use a spectral density completed at high frequency by a form factor, with
$\int_0^\infty J(\omega)\,\dd\omega<\infty$.
Its cubic form~\eqref{eq:cubic-J}, derived in the Appendix, is used only locally.

In the Wigner--Weisskopf approximation, the amplitude changes slowly over
the memory time of the field. We replace $a(t-s)$ by $a(t)$ and extend the
upper limit of the memory integral to infinity. We use ``reduced'' for quantities obtained in this approximation.
Denote the reduced amplitude by $a_{\rm red}$. It obeys
\begin{equation}\label{eq:WW}
 \begin{gathered}
 \dot a_{\rm red}=-\left(\frac{\Gamma_0}{2}+\ii\delta_{\rm rad}\right)a_{\rm red},\\
 \Gamma_0=2\pi J(\ww),\qquad
 \delta_{\rm rad}=\mathcal P\int_0^\infty
                     \frac{J(\omega)}{\ww-\omega}\,\dd\omega.
 \end{gathered}
\end{equation}
Here $\mathcal P$ stands for the Cauchy principal value.
The quantity $\delta_{\rm rad}$ is a radiative shift in angular-frequency units.
It affects the phase. Write $\xr=|a_{\rm red}|^2$ for the upper probability in the reduced
model, distinguished from the exact probability $x$. In the ordinary model,
$\xr(t)=\ee^{-\Gamma_0t}$. For the local cubic density~\eqref{eq:cubic-J},
\begin{equation}\label{eq:ordinary-result}
 \Gamma_0=\frac{\mut^2\ww^3}{3\pi\hbar\cl^3},\qquad
 \tau_0=\Gamma_0^{-1}
       =\frac{3\pi\hbar^4\cl^3}{\mut^2(\DelE)^3}.
\end{equation}

The comparison with the quartic model below is made within the Wigner--Weisskopf
approximation and with the same reservoir.

The mean survival time, which will also be used for a nonexponential law,
is defined by
\begin{equation}\label{eq:mean-definition}
 \tau_{\rm mean}[x]=\int_0^\infty x(t)\,\dd t.
\end{equation}
It equals $\tau_0$ for the reduced exponential solution. In the
one-excitation sector it integrates the upper electronic probability.

\section{Quantum mechanics from classical mechanics. Deformation of quantum mechanics.}\label{sec:q-model}

It was shown in~\cite{NP,B} that deterministic theories generalizing quantum mechanics can be obtained from classical mechanics with a restricted set of observables (we assume that our devices can measure only a part of the observables).  Considering classical mechanics on complex projective space and defining ``observable observables'' as quadratic expressions, we obtain conventional quantum mechanics. (This construction is closely related to Weinberg's nonlinear quantum mechanics.) Another particular case was described in~\cite{S}, where it was proved that quantum mechanics can be considered as classical mechanics near equilibrium. (We assume that only observables that are quadratic with respect to coordinates and momenta can be measured and that the sum of squares of coordinates and momenta is an integral of motion. We should fix some value of this integral of motion.)

It is convenient to work in complex coordinates $Z_A, \bar Z_A$. Then the integral of motion can be written in the form
$\Ntot=\sum_A\bar Z_AZ_A$. It is invariant with respect to the gauge transformation $Z_A\to \ee^{\ii\vartheta}Z_A$, with real $\vartheta$. One can consider $Z_A$ as coordinates in an orthonormal basis of a complex Hilbert space. Then the integral of motion $\Ntot$ is the square of the norm. To obtain quantum mechanics, we should fix $\Ntot$ and consider quadratic classical observables.
We fix $\Ntot=1$ throughout, consistently with the normalization in Section~\ref{sec:ordinary}.

Classical observables (in particular, classical Hamiltonians) should be gauge invariant.
For a classical Hamiltonian $\mathcal H$, Hamilton's equations have the form
\begin{equation}\label{eq:Hamilton}
 \ii\hbar\dot Z_A=\frac{\partial\mathcal H}{\partial\bar Z_A}.
\end{equation}
(The derivative is taken
with $Z_A$ and $\bar Z_A$ treated as independent variables.)

A gauge transformation transforms a state into
the same physical state. Therefore, a solution of Hamilton's equations of the form
$Z(t)=\ee^{-\ii\Omega t}Z_*$, with a fixed vector $Z_*$ and real frequency
$\Omega$, is physically stationary: its entire time
dependence is a gauge transformation.

``Observable observables'' are gauge-invariant quadratic expressions
$\langle Z,\widehat OZ\rangle$, where $\langle\cdot,\cdot\rangle$ is the Hilbert-space
inner product and $\widehat O$ is a Hermitian operator. For a quadratic Hamiltonian
$\mathcal H=\langle Z,\widehat HZ\rangle$,
Hamilton's equations
are precisely the Schr\"odinger equation $\ii\hbar\dot Z=\widehat HZ$, and the quadratic observables evolve according to the Heisenberg equations.  This means that, with a quadratic Hamiltonian, working only with ``observable observables'' we are dealing with quantum mechanics.

Now it is clear what we should do to slightly modify quantum mechanics: we should add to the quadratic Hamiltonian a small gauge-invariant quartic term.

The two-level system of the preceding section is described by the quadratic Hamiltonian
$E_0n_0+E_1n_1$ where $n_0=|\psi_0|^2$, $n_1=\norm{\bm\psi}^2$, with $\psi_0$ the amplitude of $\ket0$
and $\bm\psi=(\psi_1,\psi_2,\psi_3)$ the amplitudes of the triplet states $\ket i$.
We deform it by adding a small quartic term invariant with respect to rotations
\[{\cal H}_4=A_0 n_0^2+A_1n_1^2.\]
Here $A_0,A_1$ are real constants.

To obtain stationary states of the deformed Hamiltonian, we should solve the equation~\eqref{eq:Hamilton}
with $Z(t)=\ee^{-\ii\Omega t}Z_*$, that is,
\begin{equation}\label{eq:stationary}
 \begin{aligned}
 \hbar\Omega\,\psi_0&=(E_0+2A_0n_0)\psi_0,\\
 \hbar\Omega\,\psi_i&=(E_1+2A_1n_1)\psi_i\quad(i=1,2,3).
 \end{aligned}
\end{equation}

It is easy to solve these equations using the assumption that $A_0,A_1$ are small. The stationary energies are the values of the Hamiltonian on normalized stationary states, giving corrections $A_0$ and $A_1$.

In what follows we take $A_0=0$, so that only the metastable level is deformed:
$\mathcal H_4=A_1n_1^2$, and the stationary atomic energies are $E_0$ and $E_1+A_1$.

In the rotating-wave approximation, the two-level system interacting with the electromagnetic field is described by a quadratic Hamiltonian
depending, in the one-excitation sector, on $\bm\psi$ and photonic variables
$\phi_\alpha$ in~\eqref{eq:state}. (Recall that $\psi_i$, $i=1,2,3$, are the amplitudes of the
three $J=1$ metastable states with no photon, and $\phi_\alpha$ is the amplitude of the
$J=0$ ground state together with one photon in mode $\alpha$.)

This quadratic Hamiltonian has the form
\begin{equation}\label{eq:H2}
 \begin{split}
 \mathcal H_2={}&E_1n_1+
       \sum_\alpha(E_0+\hbar\omega_\alpha)|\phi_\alpha|^2\\
       &+\sum_{i=1}^{3}\sum_\alpha
        (g_{i\alpha}\bar\phi_\alpha\psi_i+
         \bar g_{i\alpha}\bar\psi_i\phi_\alpha).
 \end{split}
\end{equation}

Here $n_1=\sum_{i=1}^{3}|\psi_i|^2$, as in Section~\ref{sec:ordinary}.

Hamilton's equations of $\mathcal H_2$ are
\begin{align*}
 \ii\hbar\dot\psi_i
 &=\frac{\partial\mathcal H_2}{\partial\bar\psi_i}
   =E_1\psi_i+\sum_\alpha\bar g_{i\alpha}\phi_\alpha,\\
 \ii\hbar\dot\phi_\alpha
 &=\frac{\partial\mathcal H_2}{\partial\bar\phi_\alpha}
   =(E_0+\hbar\omega_\alpha)\phi_\alpha
     +\sum_{i=1}^{3}g_{i\alpha}\psi_i.
\end{align*}
These are exactly the Schr\"odinger equations~\eqref{eq:ordinary-psi}--\eqref{eq:ordinary-phi}.

To obtain a Hamiltonian describing the interaction of the two-level system with photons in the modified quantum theory, we add $\mathcal H_4$:
\begin{equation}\label{eq:H4}
 \mathcal H=\mathcal H_2+\mathcal H_4.
\end{equation}
 This is the chosen quartic deformation. Notice that we did not deform the Hamiltonian describing photons, and did not deform the magnetic moment $\mut$.

 \section {Decay in the modified quantum theory}\label{sec:quartic}

We now repeat the calculation of Section~\ref{sec:ordinary} for the
Hamiltonian~\eqref{eq:H4}.

Hamilton's equations are
\begin{align}
 \ii\hbar\dot\psi_i&=(E_1+2A_1n_1)\psi_i
                    +\sum_\alpha\bar g_{i\alpha}\phi_\alpha,\label{eq:nonlinear-psi}\\
 \ii\hbar\dot\phi_\alpha&=(E_0+\hbar\omega_\alpha)\phi_\alpha
                    +\sum_{i=1}^{3} g_{i\alpha}\psi_i.\label{eq:nonlinear-phi}
\end{align}
They coincide with~\eqref{eq:ordinary-psi}--\eqref{eq:ordinary-phi}, except for the
diagonal term $2A_1n_1$ in~\eqref{eq:nonlinear-psi}, coming from
$\partial\mathcal H_4/\partial\bar\psi_i=2A_1n_1\psi_i$.

After substituting~\eqref{eq:freephase} into
\eqref{eq:nonlinear-psi}--\eqref{eq:nonlinear-phi}, we obtain the equations for $a_i$ and
$b_\alpha$:
\begin{align}
 \dot a_i(t)&=-\frac{2\ii}{\hbar}A_1n_1(t)\,a_i(t)-\frac{\ii}{\hbar}\sum_\alpha
       \bar g_{i\alpha}b_\alpha(t)\ee^{\ii(\ww-\omega_\alpha)t},\label{eq:quartic-a}\\
 \dot b_\alpha(t)&=-\frac{\ii}{\hbar}\sum_{j=1}^{3}
       g_{j\alpha}a_j(t)\ee^{-\ii(\ww-\omega_\alpha)t}.\label{eq:quartic-b}
\end{align}
To remove the additional diagonal term in~\eqref{eq:quartic-a} directly, define
\begin{equation}\label{eq:beta}
 \theta(t)=\frac{2A_1}{\hbar}\int_0^t n_1(s)\,\dd s,
\end{equation}
and set
\begin{equation}\label{eq:transformed-amplitudes}
 a_i(t)=\ee^{-\ii\theta(t)}\widetilde a_i(t).
\end{equation}
The photon amplitudes $b_\alpha$ are unchanged.

The new amplitudes $\widetilde a_i$ have the same moduli as $a_i$, so the sector
norms~\eqref{eq:norms} of the Appendix and $x$ are unchanged.
The amplitudes $\widetilde a_i$ and $b_\alpha$ satisfy
\begin{align}
 \dot{\widetilde a}_i(t)&=-\frac{\ii}{\hbar}\sum_\alpha
       \bar g_{i\alpha}b_\alpha(t)\ee^{\ii(\ww-\omega_\alpha)t+\ii\theta(t)},\label{eq:tilde-a}\\
 \dot b_\alpha(t)&=-\frac{\ii}{\hbar}\sum_{j=1}^{3}
       g_{j\alpha}\widetilde a_j(t)\ee^{-\ii(\ww-\omega_\alpha)t-\ii\theta(t)},\label{eq:tilde-b}
\end{align}
which are~\eqref{eq:ordinary-a}--\eqref{eq:ordinary-b} with the additional phase $\theta$.
Since $b_\alpha(0)=0$, integration of~\eqref{eq:tilde-b} gives
\[
 b_\alpha(t)=-\frac{\ii}{\hbar}\sum_{j=1}^{3}g_{j\alpha}
       \int_0^t \widetilde a_j(t')\ee^{-\ii(\ww-\omega_\alpha)t'-\ii\theta(t')}\,\dd t'.
\]
Substitution into~\eqref{eq:tilde-a} yields
\begin{equation}\label{eq:nonlinear-matrix}
 \dot{\widetilde a}_i(t)=-\sum_{j=1}^{3}\int_0^tK_{ij}(t-t')
 \ee^{\ii[\theta(t)-\theta(t')]}\widetilde a_j(t')\,\dd t',
\end{equation}
with the kernel $K_{ij}$ of~\eqref{eq:memory-matrix}.
The nonlinear part of the phase is independent of the photon mode $\alpha$.
Rotational invariance again makes the kernel proportional to $\delta_{ij}$. A fixed upper polarization is preserved.
Writing $\widetilde a_i(t)=u_i \widetilde a(t)$, the scalar memory equation is
\begin{equation}\label{eq:nonlinear-memory}
 \dot{\widetilde a}(t)=-\int_0^tK(s)\ee^{\ii[\theta(t)-\theta(t-s)]}
       \widetilde a(t-s)\,\dd s,\qquad \widetilde a(0)=1.
\end{equation}
Here $x=|\widetilde a|^2$ and $K$ is the same kernel as in~\eqref{eq:J}.
This is the central reduction of the modified decay problem: the ordinary kernel is multiplied by the population-dependent phase.
This equation is exact in the specified rotating-wave model.
For $A_1=0$, $\theta=0$ and it reduces to~\eqref{eq:ordinary-memory}.

\subsection{Nonlinear Wigner--Weisskopf approximation}\label{sec:q-rate}

The phase $(\ww-\omega_\alpha)t+\theta(t)$ in the coupling has derivative
$\wtr(x(t))-\omega_\alpha$, where
\begin{equation}\label{eq:wtr}
 \wtr(x)=\ww+\frac{2A_1x}{\hbar}.
\end{equation}
Thus $\wtr(x)$ is the resonance frequency when the population $x$ is frozen
over the memory time, as is done below.
The initial resonance-energy shift $2A_1$ is twice the stationary energy
correction $A_1$, because the equations of motion involve the derivative
$2A_1n_1$ of $A_1n_1^2$. The radiative shift is treated separately.

In the short-memory regime, $\widetilde a$ and $x$ change little over the memory
time of the field. We freeze the population $x$ at its current value in the phase and
make the same replacement of the amplitude as in Section~\ref{sec:ordinary}.
The ordinary calculation then applies locally in time, at the frequency
$\wtr(x)$. Define the rate as a function of the population by
\begin{equation}\label{eq:rate-function}
 \Gamma(x)=2\pi J(\wtr(x)),\qquad \Gamma_0=\Gamma(0),
\end{equation}
and let $\delta_{\rm rad}(x)$ be the radiative shift of~\eqref{eq:WW} with $\ww$
replaced by $\wtr(x)$. The reduced amplitude $\widetilde a_{\rm red}$, with upper probability
$\xr=|\widetilde a_{\rm red}|^2$, obeys
\begin{equation}\label{eq:nonlinear-WW}
 \dot{\widetilde a}_{\rm red}=-\left(\frac{\Gamma(\xr)}{2}+\ii\delta_{\rm rad}(\xr)\right)\widetilde a_{\rm red}.
\end{equation}

Taking the derivative of $\xr=|\widetilde a_{\rm red}|^2$, we obtain
\begin{equation}\label{eq:reduced}
 \dot x_{\rm red}=-\Gamma(\xr)\xr,\qquad \xr(0)=1.
\end{equation}
The short-memory conditions for this step are discussed in the Appendix.

Consider the local cubic law~\eqref{eq:cubic-J} and put
\begin{equation}\label{eq:smallparameter}
 \varepsilon=\frac{|A_1|}{\DelE}\ll1.
\end{equation}
To first order in the quartic coefficient,
\begin{equation}\label{eq:first-rate}
 \frac{\Gamma(x)}{\Gamma_0}
 =1+\frac{6A_1x}{\DelE}+O(\varepsilon^2).
\end{equation}
On bounded intervals of the scaled time $\Gamma_0t$, integration of the
reduced law~\eqref{eq:reduced} gives
\begin{equation}\label{eq:first-survival}
 \begin{split}
 \xr(t)=\ee^{-\Gamma_0t}\left[1-\frac{6A_1}{\DelE}
   (1-\ee^{-\Gamma_0t})\right]+O(\varepsilon^2).
 \end{split}
\end{equation}
The term proportional to $A_1$ changes the shape of the curve.

\subsection{Mean lifetime and decay curve}\label{sec:q-lifetime}

Since the reduced decay is generally not exponential, there is no unique
definition of a lifetime. A natural choice is the mean survival time.
When $\Gamma(x)>0$ for $0\le x\le1$, substituting
$\dd t=-\dd x/[\Gamma(x)x]$ in~\eqref{eq:mean-definition} gives
\begin{equation}\label{eq:reduced-mean}
 \tau_{\rm red}=\tau_{\rm mean}[\xr]
       =\int_0^1\frac{\dd x}{\Gamma(x)}.
\end{equation}
For the cubic law, expansion of the integrand using~\eqref{eq:first-rate} yields
\begin{equation}\label{eq:mean-result}
 \frac{\tau_{\rm red}}{\tau_0}
       =1-\frac{3A_1}{\DelE}+O(\varepsilon^2).
\end{equation}
This is the correction relative to the ordinary model with the same
gap $\DelE$, matrix element and reservoir.
Equivalently, the absolute change of the mean is
${\tau_{\rm red}-\tau_0=-3\tau_0A_1/\DelE+O(\tau_0\varepsilon^2)}$.

For a fixed spectral density $J$ that is twice continuously differentiable
near $\ww$, with $J(\ww)>0$, Taylor expansion more generally gives
\begin{equation}\label{eq:general-expansion}
 \begin{gathered}
 \Gamma(x)=\Gamma_0+\Gamma'(0)x+O(\Gamma_0\varepsilon^2),\\
 \frac{\tau_{\rm red}}{\tau_0}
 =1-\frac{\Gamma'(0)}{2\Gamma_0}+O(\varepsilon^2).
 \end{gathered}
\end{equation}
Here and below, a prime on $\Gamma$ denotes a population derivative, with
$\Gamma'(0)=\allowbreak(2A_1/\hbar)\allowbreak(2\pi J)'(\ww)$.
The constants in these remainders may depend on the fixed reservoir.

The reference gap matters. The stationary atomic energies of
Section~\ref{sec:q-model} have difference $\Ec=\DelE+A_1$.
An ordinary atom with this corrected gap and the same coupling has rate
$\Gamma_c=\Gamma(1/2)$. Equation~\eqref{eq:general-expansion} implies
$\Gamma_c\tau_{\rm red}=1+O(\varepsilon^2)$.
Thus the mean agrees with that ordinary lifetime to first order, while
the cubic reduced curve~\eqref{eq:first-survival} changes shape at first order.
This comparison uses $\Ec$, whereas~\eqref{eq:mean-result} uses $\DelE$.

\paragraph{Variation of the decay rate.}
The logarithmic decay rate of the reduced solution is
\begin{equation}\label{eq:lograte}
 r(t)=-\frac{\dd}{\dd t}\log\xr(t)=\Gamma(\xr(t)).
\end{equation}
Its initial and late values are
\begin{align}
 \Gamma_{\rm initial}&=\Gamma(1)
       =\Gamma_0\left(1+\frac{2A_1}{\DelE}\right)^3,
       \label{eq:initial}\\
 \Gamma_{\rm late}&=\Gamma(0)=\Gamma_0,\label{eq:late}
\end{align}
where the displayed expression for $\Gamma(1)$ uses the cubic law.
Both rates belong to the reduced description (see the Appendix).
The initial effective lifetime is consequently
$\Gamma_{\rm initial}^{-1}=\tau_0[1-6A_1/\DelE+O(\varepsilon^2)]$.
The different initial and late logarithmic rates show how the curve departs
from an exponential on the decay timescale.

Expanding $\Gamma(1)$ and the mean in~\eqref{eq:general-expansion} gives
\begin{equation}\label{eq:initial-mean-relation}
 \frac{\Gamma_{\rm initial}}{\Gamma_{\rm late}}-1
 =-2\left(\Gamma_{\rm late}\tau_{\rm red}-1\right)+O(\varepsilon^2).
\end{equation}
If $J'(\ww)=0$, both linear terms in~\eqref{eq:initial-mean-relation} vanish.
This relation does not require the cubic law or an independent value of
$\Gamma_0$. It compares quantities from a single reduced decay curve.
For observations much shorter than the lifetime, the late rate and the mean
must be inferred by extrapolation. The achievable precision then depends on
the observation time, photon statistics and control of ordinary losses.

Differentiating~\eqref{eq:lograte} and using~\eqref{eq:reduced} gives
\begin{equation}\label{eq:curvature}
 \frac{\dd^2}{\dd t^2}\log\xr(t)
 =\Gamma'(\xr(t))\Gamma(\xr(t))\xr(t).
\end{equation}
For the cubic rate, $\log\xr$ is convex for $A_1>0$ and concave for $A_1<0$.
More generally, through the frequency dependence of $J$, the population-dependent resonance frequency~\eqref{eq:wtr} makes the reduced decay nonexponential whenever $\Gamma$ is positive and nonconstant on $[0,1]$.
Nonexponentiality of a measured survival does not by itself distinguish
a deformation from ordinary loss processes.

\section{Conclusion}\label{sec:conclusion-proposed}

We calculated the change in the mean lifetime of metastable helium in the case when quantum mechanics is slightly modified. We proved that in our approximation, for the cubic law~\eqref{eq:cubic-J} and to first order in $A_1/\DelE$, this change is
$-3\tau_0A_1/\DelE$, where $\tau_0$ is the ordinary lifetime at the same gap $\DelE$. Taking into account that the lifetime $\tau_0$ is about two hours, we conclude that the change can be significant
even if $A_1/\DelE$ (the relative change of the energy level with respect to the gap between the excited state and ground state) is small. 

A long lifetime enlarges this absolute change, but not the fractional
mean-lifetime change $3|A_1|/\DelE$ at first order.
At the corrected gap $\Ec=\DelE+A_1$ and with the same coupling, the mean
agrees with the ordinary lifetime to first order, whereas the cubic decay
curve changes shape at first order.

Metastable helium, whose ordinary radiative lifetime was calculated
from first principles, provides a system in which these predictions can
be compared with experiment. 

We also analyzed the deviation from the exponential law of decay in modified quantum mechanics on the example of metastable helium. It is important to emphasize that our calculations can be generalized to other decay models with a specified deformation~\cite{NS}. They show that in the situation when the lifetime cannot be calculated theoretically, one can hope to discover a violation of the laws of QM by looking at the deviation from the exponential law.

\section*{Acknowledgements}

We are indebted to L.Alvarez-Gaume, M. Luty, N.Nekrasov, J.Terning, and A. Vainshtein for useful discussions.

AI (ChatGPT) was helpful in the preparation of this paper.

\section*{Appendix}
\setcounter{equation}{0}
\renewcommand{\theequation}{A\arabic{equation}}
\renewcommand{\theHequation}{A\arabic{equation}}

\paragraph{Field and coupling conventions.}

In a normalization volume $V$, use two transverse real orthonormal polarization vectors
$\bm\epsilon_{\mathbf k\lambda}$, with $\lambda=1,2$, for each wave vector
$\mathbf k$. The photon angular frequency is
$\omega_{\mathbf k}=\cl|\mathbf k|$, where $\cl$ is the speed of light.
We use rationalized Heaviside--Lorentz electromagnetic units, with $\cl$ explicit.
Each sum over $\mathbf k$
runs over the nonzero wave vectors allowed in the normalization volume.
Let $\hat a_{\mathbf k\lambda}$ and $\hat a^\dagger_{\mathbf k\lambda}$ be
the photon annihilation and creation operators. These operators satisfy
$[\hat a_{\mathbf k\lambda},\hat a^\dagger_{\mathbf k'\lambda'}]
=\delta_{\mathbf k\mathbf k'}\delta_{\lambda\lambda'}$. The one-photon state is $\ket{1_\alpha}=\hat a^\dagger_\alpha\ket{0_\gamma}$.

In Coulomb gauge,
\begin{equation}\label{eq:B}
 \mathbf B(\mathbf0)=\ii\sum_{\mathbf k}\sum_{\lambda=1}^{2}
 \cl\sqrt{\frac{\hbar}{2\omega_{\mathbf k}V}}
 (\mathbf k\times\bm\epsilon_{\mathbf k\lambda})
 (\hat a_{\mathbf k\lambda}-\hat a^\dagger_{\mathbf k\lambda}).
\end{equation}

With $\mathbf e_i$ the unit vector along axis $i$, the
coupling~\eqref{eq:g} satisfies
\begin{equation}\label{eq:g2}
 |g_{i,\mathbf k\lambda}|^2=
 \mut^2\frac{\hbar\cl^2}{2\omega_{\mathbf k}V}
 |\mathbf e_i\cdot(\mathbf k\times\bm\epsilon_{\mathbf k\lambda})|^2.
\end{equation}

In the strict nonrelativistic Russell--Saunders, or $LS$-coupling, limit, the
transition is forbidden. Relativistic and spin-dependent corrections generate
the small physical matrix element~\cite{LP}.

\paragraph{Rotating-wave model.}

In the ordinary weak-coupling calculation, the omitted counter-rotating
terms create only off-resonant virtual states at leading order. They mainly
affect energy renormalization and higher-order corrections and do not modify
the leading on-shell (energy-conserving) decay rate.
Writing $\mathbf B=\mathbf B^{(+)}+\mathbf B^{(-)}$, where the first term
contains annihilation operators and the second their adjoints, the retained
interaction is
\begin{equation}\label{eq:VRWA}
 V_{\rm RWA}=-\mut\sum_{i=1}^{3}
 (\ket0\bra i\,B_i^{(-)}+\ket i\bra0\,B_i^{(+)}).
\end{equation}
Define the projectors $P_0=\ket0\bra0$ and
$P_1=\sum_{i=1}^{3}\ket i\bra i$. The free Hamiltonian is
$H_0=E_0P_0+E_1P_1+H_\gamma$, where
$H_\gamma=\sum_\alpha\hbar\omega_\alpha\hat a^\dagger_\alpha\hat a_\alpha$.
Tensor identities on the atomic and photon factors are suppressed.
The operator $Q=P_1+\sum_\alpha\hat a^\dagger_\alpha\hat a_\alpha$  commutes with $H_0+V_{\rm RWA}$.

Define the squared sector norms and their conserved sum by
\begin{equation}\label{eq:norms}
 n_1=\sum_{i=1}^{3}|\psi_i|^2,\qquad n_0=\sum_\alpha|\phi_\alpha|^2,
 \qquad n_1+n_0=1.
\end{equation}

In the variables of~\eqref{eq:state}, the Schr\"odinger equation
$\ii\hbar\dot Z=(H_0+V_{\rm RWA})Z$ reads
\begin{align}
 \ii\hbar\dot\psi_i&=E_1\psi_i
                    +\sum_\alpha\bar g_{i\alpha}\phi_\alpha,\label{eq:ordinary-psi}\\
 \ii\hbar\dot\phi_\alpha&=(E_0+\hbar\omega_\alpha)\phi_\alpha
                    +\sum_{i=1}^{3} g_{i\alpha}\psi_i.\label{eq:ordinary-phi}
\end{align}

\paragraph{Local cubic spectral density.}

For the point coupling~\eqref{eq:Hint} with constant $\mut$, $J(\omega)$ is cubic. Writing $k=|\mathbf k|$, we obtain
\begin{align}
 J(\omega)&=\frac{\mut^2\cl^2}{2\hbar}
       \frac{8\pi}{3(2\pi)^3}
       \int_0^\infty\frac{k^4}{\cl k}\delta(\omega-\cl k)\,\dd k
       =\frac{\mut^2}{6\pi^2\hbar\cl^3}\omega^3.\label{eq:cubic-J}
\end{align}
For a realistic atom, this expression is multiplied by an atomic form factor
$|F(\omega)|^2$.
The cubic expression is used
locally, where the effective matrix element and $F$ may be treated as constant.
The value of the measured rate fixes a normalization, but does not determine
the local slope of the effective matrix element~\cite{LP}.

Formula~\eqref{eq:ordinary-result} also follows from Fermi's golden rule. In this approximation,
the natural linewidth is $\Gamma_0$, the full width at half maximum of the
emission line in angular-frequency units.

\paragraph{Validity of the reduced description.}

For metastable helium, $\ww\simeq3.0\times10^{16}$ s$^{-1}$ and the
reported lifetime is about $7.87\times10^3$ s~\cite{H}. These scales give
$\Gamma_0/\ww\simeq4.2\times10^{-21}$.
The upper-state amplitude changes on a scale $1/\Gamma_0$, while the
field kernel changes on microscopic optical/atomic scales in the short-memory
regime. The separation is about twenty orders of magnitude.
With a form factor smooth on atomic frequency scales, $J(\omega)$ is
extremely smooth over the resonance width $|\omega-\ww|\lesssim\Gamma_0$,
so $J(\omega)\simeq J(\ww)$ across this narrow interval.
A bound on the memory error also requires information about the
completed spectral density. No quantitative memory-error bound for helium is inferred
from the ratio $\Gamma_0/\ww$ alone.

The exact memory equation~\eqref{eq:ordinary-memory} is not an exponential law even in ordinary
quantum mechanics. For an integrable spectral density, $\dot a(0)=0$,
and the survival initially has a quadratic rather than a linear decrease.
Its behavior at very long times also depends on the continuum spectrum.
The exponential approximation describes the intermediate decay regime.

The reduced law~\eqref{eq:reduced} assumes a regular reservoir with a sufficiently short memory
and a slowly varying transition frequency over that memory time.
The cubic approximation for $J$ must hold throughout the interval swept by $\wtr(x)$, $0\le x\le1$, whenever that approximation is used.

The rates~\eqref{eq:initial}--\eqref{eq:late} are rates of the reduced law. The exact initial derivative of
the survival is zero, as in the ordinary memory equation~\eqref{eq:ordinary-memory}. The late rate~\eqref{eq:late}
is also distinct from the ultimate continuum tail.

Within the reduced description, the measurement of Ref.~\cite{H}, which
observes the XUV emission over windows of about $0.5$~s, much shorter than
$\tau_0$, probes the initial rate~\eqref{eq:initial}, provided that $\xr$
remains close to $1$, the memory transient has passed and the calibration is
unchanged. The lifetime inferred from it then corresponds to
$\Gamma_{\rm initial}^{-1}$ rather than directly to the mean survival time
$\tau_{\rm red}$.

The first-order truncation of the reduced law~\eqref{eq:reduced} is
$\dot x_{\rm red}=-\Gamma_0\xr-\Gamma'(0)\xr^2$.
For ${\Gamma'(0)\geq0}$, this is also an ordinary population-loss law with
constant one- and two-body loss coefficients. Such losses satisfy
relation~\eqref{eq:initial-mean-relation} to first order in
$\Gamma'(0)/\Gamma_0$. The relation is therefore an internal consistency
check, not a unique signature of a deformation.

\paragraph{Comparison with Weinberg's nonlinear QM.}\hypertarget{weinberg-appendix}{}
Throughout, we used the quartic Hamiltonian~\eqref{eq:H4} on the unit norm sphere
to deform the two-level QM.
In Weinberg's framework~\cite{W} a classical Hamiltonian is a homogeneous function
of degree one in $Z$ and in $\bar Z$. Two-level atoms in this framework were considered in~\cite{Cz}.

 One can check that our quartic deformation leads to results that are essentially equivalent to deformations constructed in~\cite{Cz}.
This comparison concerns the specified atom--field model. It does not
select prescriptions for systems of several atoms, statistical mixtures or
measurements. General questions of observables and probabilities are
discussed in Chapter~4 of~\cite{B}.

\paragraph{Experimental considerations.}

A precise measurement of the lifetime and decay law of metastable helium is a promising way to find deviations from QM. It also
could provide a complementary test of nonlinear QM alongside existing
precision tests~\cite{Bo,CH,Wa,Ma}, which search for population- or
polarization-dependent frequency shifts.
Current XUV detectors can detect individual emitted photons; hence one can hope to observe the number of decays with high precision. A difficult problem is to take into account the losses due to collisions with other atoms and
with the boundary.
\end{document}